\documentclass[twocolumn,amsmath,amssymb,aps,pra,superscriptaddress,floatfix]{revtex4-2}

\usepackage{graphicx}
\usepackage{dcolumn}
\usepackage{bm}
\usepackage{physics}
\usepackage{color}
\usepackage{booktabs}
\usepackage{microtype}
\usepackage{tikz}
\usetikzlibrary{arrows.meta,calc,positioning,fit,backgrounds}
\usepackage{amsmath,amssymb}
\usepackage{graphicx}
\usepackage{hyperref}
\usetikzlibrary{arrows.meta,positioning,calc,fit,decorations.pathreplacing}

\usetikzlibrary{shapes,arrows.meta,positioning,decorations.pathmorphing,fit}
\usepackage{pgfplots}
\pgfplotsset{compat=1.18}

\definecolor{panelA}{RGB}{31, 119, 180}   
\definecolor{panelB}{RGB}{255, 127, 14}   
\definecolor{panelC}{RGB}{44, 160, 44}    
\definecolor{panelD}{RGB}{214, 39, 40}    
\definecolor{gridgray}{RGB}{230, 230, 230}

\usetikzlibrary{shapes,arrows.meta,positioning,decorations.pathmorphing}
\hypersetup{
    colorlinks=true,
    linkcolor=blue,
    citecolor=blue,
    urlcolor=blue
}

\begin{document}

\title{Helium  States  Inspired by Quantum Photonics Circuit}

\author{F. Mirmasoudi}
\author{B. Ghavami}
\email{badie.ghavami@ipm.ir}
\affiliation{Iranian Center for Quantum Technologies (ICQTs), Tehran, Iran}


\date{\today}
\begin{abstract}
We present a four-mode Fock-state photonics implementation of the variational quantum deflation (VQD) algorithm to calculate the ground-state and excited-state energies for the neutral helium atom in a CAS(2,2) Subspace. Its CAS(2,2) electronic Hilbert space is encoded into a bosonic Fock-space representation and subsequently optimized using a continuous-variable quantum circuit.  The circuit architecture under study is constructed of single-mode phase shifts, beam splitters, and cross-Kerr interactions. The variational circuit is initialized in the Fock state $\ket{1100}$ and its performance is benchmarked against exact diagonalization of the active-space configuration-interaction (CASCI) Hamiltonian. We also extend the method to optimize from the first to three excited states of the helium atom with the aid of the variational quantum deflation $(VQD)$ method in a CAS(2,2). We then investigate the performance of this circuit architecture as a basis for implementing the VQD method. In the ideal noiseless simulation setting, we see that the survival probability of the ground and the excited states within the target $CAS$ subspace is unity, because the total photon number and the photon number in the two spin sectors $\alpha$ and $\beta$ are preserved. We demonstrate that the proposed approach enables the determination of the ground and excited-state energies within the CAS(2,2) electronic Hilbert space using a continuous-variable quantum circuit and the variational quantum deflation ($VQD$) algorithm, while satisfying the orthogonality conditions with negligible errors.

\end{abstract}
\maketitle

\section{Introduction}
\label{sec:introduction}

Describing atomic or molecular electronic states at their high-energy levels remains a challenging problem in the field of quantum chemistry for systems which possess an electronic ground state that is composed of several contributing electronic states~\cite{mcardle2020quantum,cao2019quantum}. Electronic excited states are important in processes like optical excitation, photochemical reactions, charge transfer, and dynamics that are not adiabatic (i.e., the electronic and nuclear motions are not coupled). There are several methods of electronic structure representation that have been used for describing excited states, however the performance of the classical electronic structure methods is limited by strong multi-reference character or by the existence of nearly degenerate electronic states. Therefore, many methods of electronic structure calculation that are based on Configuration Interaction with Singles and Doubles (CISD) or Equation of Motion Coupled-Cluster (EOM-CC), and Time-Dependent Density Functional Theory (TDDFT) require a large amount of computer power as additional active orbitals become important in the calculations~\cite{szabo1996modern,helgaker2014molecular,bartlett2007coupled,dreuw2015algebraic,shepard1987multireference}.

The advent of Variational Quantum Algorithms (VQAs) executed within a hybrid or quantum-classical setting has led to them becoming the leading candidates for realizing a useful quantum computation on the noisy intermediate-scale quantum (NISQ) hardware~\cite{cerezo2021variational,peruzzo2014variational,bharti2022noisy,mcclean2016theory,peruzzo2014variational}. Among these variational quantum algorithms, the Variational Quantum Eigensolver (VQE) has been very effective in finding the potential energy surface of the ground states of molecules.Variational approaches have also been investigated for quantum dynamics and ground-state problems using neural quantum states, providing a complementary perspective on variational quantum simulation~\cite{EmpiricalStudy2023}.  The application of the variational methods to the electronic excited states of molecules is fundamentally challenging because the standard Rayleigh-Ritz minimization will converge to the global ground state in all cases where no specific orthogonality conditions are applied. Several methods have been developed, including quantum subspace expansion (QSE)~\cite{mcclean2017hybrid} and subspace-search VQE (SS-VQE)~\cite{nakanishi2019subspace} as well as a number of other folded spectrum (FS) techniques~\cite{colless2018computation,wang2019quantum,parrish2019quantum,jones2019variational}.
While there are several ways to calculate the exact electronic states of an atom or molecule, it has been shown that the Variational Quantum Deflation (VQD) algorithm~\cite{higgott2019variational} is 
one of the most prominent approaches among all the variations of sequential orthogonalized circuits. The VQD algorithm works by applying dynamic deflation potentials to penalize overlaps with previously converged lower-energy states, thereby maintaining the variational bound without increasing the register footprint of the spectral components. VQD protocols are not limited to excited-state energetic, it used to extract transition amplitudes and off-diagonal spectral properties ~\cite{kottmann2022calculating}.

To date, the predominant implementation of VQAs has relied on gate-based qubit 
platforms (e.g., superconducting circuits and trapped-ion chains)~\cite{kandala2017hardware,hempel2018quantum,omalley2016scalable,nam2020ground}. 
In these architectures, mapping fermionic Hamiltonians via Jordan-Wigner or Bravyi-Kitaev 
transformations typically yields non-local Pauli strings that demand deep circuits, ancilla 
overhead, and explicit software-level penalty terms to maintain global spatial and spin symmetries. 
Conversely, photonic continuous-variable (CV) and discrete Fock-state platforms offer a compelling 
physical alternative~\cite{bourassa2021blueprint,arrazola2021quantum,kok2007linear,aspuru2012photonic,carolan2015universal,bromley2020applications,sparrow2018simulating,killoran2019continuous,aspuru2012photonic}.  Morover,
Continuous-variable quantum information has also been studied in photonic settings using nonclassical optical states, although for tasks distinct from electronic-structure simulation ~\cite{TemelSkoric2025}. By naturally encoding quantum 
information into optical spatial or temporal modes, photonics provides room-temperature operability, 
ultra-low decoherence, and most crucially exact intrinsic conservation of particle number 
$\langle \hat{N} \rangle$ and spin projection $S_z$ through unitary, beam-splitter-mediated linear 
optical transformations.

In our preceding investigation~\cite{ghavami2025variational}, we demonstrated the feasibility 
of determining the singlet ground state ($1\,^1S$) of the neutral helium atom ($E_0 \approx -2.8703\,\text{Ha}$) 
on a 4-mode integrated photonic processor. However, resolving the full low-lying manifold—including 
orthogonality among singlet and triplet manifolds within an active space—remains an open computational 
challenge for all-photonic architectures, primarily due to the stringent requirements on multi-parameter 
variational expressibility and nonlinear phase modulation.

In this work, we investigate a number-preserving continuous-variable variational quantum deflation (VQD) framework for characterizing the electronic energy spectrum of helium within the CAS(2,2) active space. The approach employs a four-mode photonic Fock-space architecture and an 18-parameter variational ansatz constructed from phase rotations, beam-splitter interactions, and an effective cross-Kerr nonlinearity. By preserving the total particle number and the spin-resolved occupations, the circuit remains confined to the intended two-particle subspace throughout the optimization, eliminating the need for post-selection to correct particle-number leakage. The VQD procedure is formulated through a projected Rayleigh quotient over the four-dimensional CAS determinant basis, with sequential deflation penalties used to obtain orthogonal approximations to the ground and excited states. Numerical simulations in Strawberry Fields show that the ground-state energy can be reproduced with high accuracy relative to the CAS full configuration interaction benchmark. The excited-state calculations recover the overall structure of the low-lying spectrum, although their accuracy is more sensitive to the expressibility of the ansatz and the non-convex optimization landscape associated with the deflation terms. These results illustrate both the potential and the current limitations of symmetry-preserving continuous-variable circuits for variational quantum chemistry.\\
The remainder of this manuscript is organized as follows: Section~\ref{sec:methodology} delineates the 
molecular CAS(2,2) Hamiltonian mapping, the 18-parameter number-conserving circuit architecture, and 
the deflation formalisms. Section~\ref{sec:results} provides a rigorous numerical dissection of the energy 
convergence profiles, overlap matrices, and penalty parameter resilience. 

\begin{figure*}[t]
\centering
\label{fig:1}
\resizebox{0.85\textwidth}{!}{ 
\begin{tikzpicture}[
    box/.style={rectangle, draw=black!80, fill=blue!10, thick, minimum width=1.4cm, minimum height=0.8cm, align=center, font=\small},
    det/.style={rectangle, draw=black!80, fill=red!15, thick, minimum width=1.2cm, minimum height=0.7cm, align=center, font=\footnotesize},
    gate/.style={rectangle, draw=black!80, fill=green!15, thick, minimum width=1.6cm, minimum height=0.8cm, align=center, font=\footnotesize},
    ansatzbox/.style={rectangle, draw=black!80, fill=gray!5, thick, minimum width=2.0cm, minimum height=2.8cm, align=center, font=\small\bfseries},
    guide/.style={thick, -{Latex[length=2.5mm]}},
    optline/.style={thick, color=blue!80!black, -{Latex[length=2mm]}},
    elecline/.style={dashed, thick, color=black!70, -{Latex[length=2mm]}}
]

    \node[font=\bfseries\sffamily] at (0.8, 2.3) {(a) TMSV Sources};
    \node[font=\bfseries\sffamily] at (4.6, 2.3) {(b) Heralding Logic};
    \node[font=\bfseries\sffamily] at (9.4, 2.3) {(c) State Injection};

    \node[box, fill=gray!20] (laser) at (-1.4, 0) {Pump\\Laser};
    \node[box] (spdc1) at (0.8, 1.1) {SPDC 1\\($\chi^{(2)}$)};
    \node[box] (spdc2) at (0.8, -1.1) {SPDC 2\\($\chi^{(2)}$)};

    \draw[optline] (laser.east) -- ++(0.4,0) |- (spdc1.west);
    \draw[optline] (laser.east) -- ++(0.4,0) |- (spdc2.west);

    \node[det] (det1) at (3.2, 1.5) {SNSPD 1\\($i_1$)};
    \node[det] (det2) at (3.2, -1.5) {SNSPD 2\\($i_2$)};

    \draw[optline] (spdc1.east) -- ++(0.3,0) |- (det1.west) node[above, pos=0.8, font=\scriptsize] {idler};
    \draw[optline] (spdc2.east) -- ++(0.3,0) |- (det2.west) node[below, pos=0.8, font=\scriptsize] {idler};

    \node[gate] (coin) at (5.0, 0) {Coincidence\\Logic ($n=1$)};
    \draw[elecline] (det1.east) -| (coin.north);
    \draw[elecline] (det2.east) -| (coin.south);

    \node[ansatzbox] (vqe) at (9.5, 0) {Ansatz\\Circuit\\$\hat{U}(\vec{\theta})$};

    \draw[optline] (spdc1.east) -- ++(0.8,0) |- (8.5, 0.8) node[pos=0.25, above, font=\scriptsize] {signal} node[left, font=\footnotesize] at (7.8, 1.0) {$q_0: \ket{1}$};
    \draw[optline] (spdc2.east) -- ++(0.8,0) |- (8.5, 0.3) node[pos=0.25, below, font=\scriptsize] {signal} node[left, font=\footnotesize] at (7.8, 0.6) {$q_1: \ket{1}$};

    \draw[optline] (7.8, -0.3) -- (8.5, -0.3) node[left, font=\footnotesize] at (7.8, -0.3) {$q_2: \ket{0}$};
    \draw[optline] (7.8, -0.8) -- (8.5, -0.8) node[left, font=\footnotesize] at (7.8, -0.8) {$q_3: \ket{0}$};

    \draw[elecline] (coin.east) -- ++(0.5,0) |- (7.2, 0.75);

    \node[right, font=\footnotesize] at (10.5, 0) {$\ket{\Psi_{\text{in}}}$};

\end{tikzpicture}
}
\caption{\textbf{Schematic of the heralded state preparation protocol.} (a) A pulsed pump laser drives two independent SPDC sources, each generating a two-mode squeezed vacuum (TMSV) state. (b) The idler modes ($i_1, i_2$) are detected by SNSPDs, triggering the heralding event ($n_{i_1}=n_{i_2}=1$). (c) Conditioned on this coincidence, signal photons are routed to rails $q_0$ and $q_1$, while $q_2$ and $q_3$ are injected with vacuum, yielding the state $\ket{\Psi_{\text{in}}} = \ket{1, 1, 0, 0}$ at the input of the ansatz circuit.}
\label{fig:hf_prep_schematic}
\end{figure*}
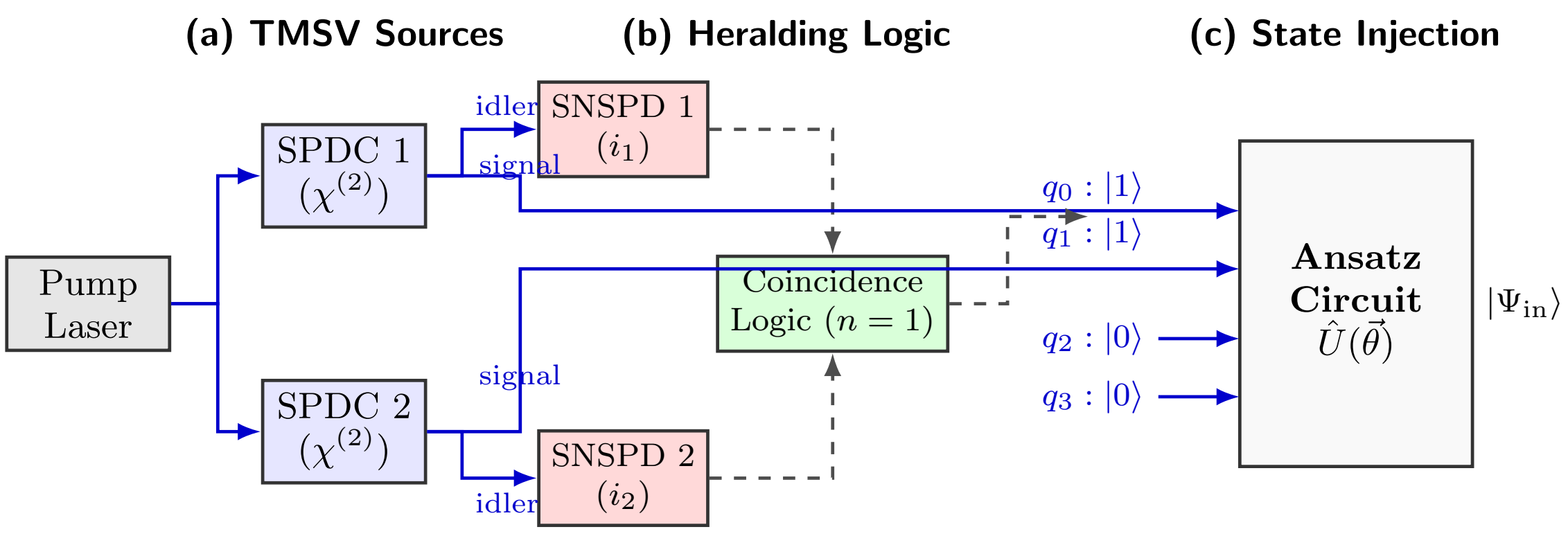

\section{Methodology and Theoretical Framework}
\label{sec:methodology}
\subsection{Electronic Hamiltonian and CAS(2,2) Active-Space Mapping}
\label{subsec:hamiltonian}

The non-relativistic electronic structure of the isolated helium atom ($Z = 2$) within the Born--Oppenheimer approximation is governed by the second-quantized fermionic Hamiltonian~\cite{helgaker2013molecular,szabo2012modern,helgaker2014molecular}:
\begin{equation}
\hat{H}_{\mathrm{elec}} = \sum_{pq} h_{pq} \hat{c}_p^\dagger \hat{c}_q + \frac{1}{2} \sum_{pqrs} g_{pqrs} \hat{c}_p^\dagger \hat{c}_r^\dagger \hat{c}_s \hat{c}_q,
\label{eq:hamiltonian}
\end{equation}
where $\hat{c}_p^\dagger$ ($\hat{c}_p$) denote standard fermionic creation (annihilation) operators satisfying the canonical anticommutation relations $\{\hat{c}_p, \hat{c}_q^\dagger\} = \delta_{pq}$ and $\{\hat{c}_p, \hat{c}_q\} = 0$. The indices $p, q, r, s$ label the spatial-spin molecular or atomic spin-orbitals $\{\chi_p(\bm{r}, \sigma) = \phi_p(\bm{r}) \otimes \chi_\sigma\}$. The scalar coefficients $h_{pq}$ and $g_{pqrs}$ represent the standard one- and two-electron integrals in physicist's notation, respectively defined as:
\begin{align}
h_{pq} &= \int \chi_p^*(\bm{x}) \left( -\frac{1}{2}\nabla^2 - \frac{Z}{|\bm{r}|} \right) \chi_q(\bm{x})\, d\bm{x}, \label{eq:one_body_integral} \\
g_{pqrs} &= \iint \frac{\chi_p^*(\bm{x}_1) \chi_r^*(\bm{x}_2) \chi_s(\bm{x}_2) \chi_q(\bm{x}_1)}{|\bm{r}_1 - \bm{r}_2|}\, d\bm{x}_1 d\bm{x}_2, \label{eq:two_body_integral}
\end{align}
where $\bm{x} \equiv (\bm{r}, \sigma)$ encapsulates spatial and spin coordinates.

To capture the essential static electron correlation effects while retaining computational tractability, the Hilbert space is truncated to a complete active space comprising two active electrons partitioned across two spatial orbitals, denoted as $\mathrm{CAS}(2,2)$~\cite{olsen2011casscf,roos1980complete}. In standard Gaussian basis sets (such as restricted Hartree--Fock orbitals expanded in 6-31G), the spatial orbitals correspond to the primary $1s$ and secondary $2s$ shells, generating four spin-orbitals: $\{1s\alpha, 1s\beta, 2s\alpha, 2s\beta\}$. Enforcing total particle-number conservation ($\hat{N}_e = 2$) and zero net spin projection ($S_z = 0$) isolates a four-dimensional invariant subspace $\mathcal{H}_{\mathrm{CAS}} \subset \mathcal{F}_{\mathrm{fermion}}$. This subspace is spanned by the orthonormal Slater determinant basis:
\begin{equation}
\mathcal{B}_{\mathrm{CAS}} = \left\{ \ket{1100}, \ket{1001}, \ket{0110}, \ket{0011} \right\},
\label{eq:cas_basis}
\end{equation}
expressed in the occupation number representation $\ket{n_{1s\alpha}, n_{1s\beta}, n_{2s\alpha}, n_{2s\beta}}$, where $\sum_p n_p = 2$ and $n_{1s\alpha} + n_{2s\alpha} = n_{1s\beta} + n_{2s\beta} = 1$.
To bridge the electronic structure with the photonic hardware, we map these four fermionic states to a four-mode bosonic Fock-space representation ($m_0, m_1, m_2, m_3$ corresponding to optical rails $q_0, q_1, q_2, q_3$). This explicit encoding translates the configuration state functions into occupation-number Fock states via:
\begin{align*}
|1\alpha 1\beta\rangle &\longleftrightarrow |1100\rangle, \\
|1\alpha 2\beta\rangle &\longleftrightarrow |1001\rangle, \\
|1\beta 2\alpha\rangle &\longleftrightarrow |0110\rangle, \\
|2\alpha 2\beta\rangle &\longleftrightarrow |0011\rangle.
\end{align*}
This mapping naturally guarantees that the total particle number and spin projections remain invariant during the optical unitary evolution.
Structurally, $\ket{1100}$ and $\ket{0011}$ represent closed-shell configurations ($|1s^2\rangle$ and $|2s^2\rangle$), whereas $\ket{1001}$ and $\ket{0110}$ span the open-shell manifold ($|1s 2s\rangle$) that gives rise to decoupled singlet ($^1\mathrm{S}$) and triplet ($^3\mathrm{S}$) electronic terms.

Projecting the full Hamiltonian $\hat{H}_{\mathrm{elec}}$ onto $\mathcal{B}_{\mathrm{CAS}}$ yields the $4 \times 4$ real-symmetric matrix $\bm{H}_{\mathrm{CAS}}$ with elements $H_{ij} = \mel{\phi_i}{\hat{H}_{\mathrm{elec}}}{\phi_j}$ ($i,j \in \{0,1,2,3\}$). Canonical spectral decomposition:
\begin{equation}
\bm{H}_{\mathrm{CAS}} \bm{v}_k = E_k \bm{v}_k \quad (k = 0, 1, 2, 3)
\label{eq:exact_diag}
\end{equation}
provides the exact ground-truth active-space eigenvalues $E_k$ and target configuration-interaction eigenvectors $\bm{v}_k$, against which our variational photonic simulations are benchmarked.

\subsection{Photonic Ansatz Architecture and Symmetry Preservation}
\label{subsec:photonic_ansatz}

The variational quantum circuit is instantiated across an $M=4$ spatial-rail optical architecture. The input state is prepared in the dual-photon Fock configuration:
\begin{equation}
\ket{\Psi_{\mathrm{in}}} = \ket{1, 1, 0, 0} \equiv \hat{a}_0^\dagger \hat{a}_1^\dagger \ket{0}^{\otimes 4},
\label{eq:input_state}
\end{equation}

which physically mirrors the non-interacting closed-shell Hartree--Fock reference state $\ket{1s\alpha, 1s\beta, 0, 0}$ within the mapped CAS(2,2) manifold. In an experiment, the input state $\ket{1,1,0,0}$ could be prepared by injecting two heralded single photons into separate spatial rails, using spontaneous parametric down-conversion (SPDC) or spontaneous four-wave mixing (SFWM) sources~\cite{eisaman2011invited, paesani2020generation,senellart2017high}. You can find  implementation of this protocol in Fig.~\ref{fig:1}. 

\definecolor{indigo}{RGB}{75,0,130}
\begin{figure*}[t!]
\centering
\label{fig:circuit_schematic}
\resizebox{\textwidth}{!}{%
\begin{tikzpicture}[
    >=latex,
    font=\sffamily,
    rail/.style={draw=black!85, line width=1.15pt},
    rgateA/.style={
        draw=blue!85!black, fill=cyan!50!blue!35, line width=0.9pt,
        minimum width=1.25cm, minimum height=0.55cm,
        rounded corners=4pt, font=\small\bfseries\itshape
    },
    rgateB/.style={
        draw=green!60!black, fill=green!65!black!30, line width=0.9pt,
        minimum width=1.25cm, minimum height=0.55cm,
        rounded corners=4pt, font=\small\bfseries\itshape
    },
    bsbox/.style={
        draw=blue!75!black, fill=blue!10!indigo!10, line width=0.9pt,
        minimum width=0.95cm, rounded corners=5pt
    },
    ckbox/.style={
        draw=orange!95!black, fill=orange!60!yellow!40, line width=0.9pt,
        minimum width=1.45cm, minimum height=0.58cm,
        rounded corners=4pt, font=\scriptsize\bfseries
    },
    sidebox/.style={
        draw=blue!75!black, fill=blue!10!indigo!10, line width=1.1pt,
        rounded corners=8pt, align=center
    }
]

\def\yA{3.45}
\def\yB{2.30}
\def\yC{1.15}
\def\yD{0.00}

\draw[draw=cyan!80!blue, fill=cyan!12, dashed, line width=1.1pt, rounded corners=8pt] 
    (0.8, -0.65) rectangle (7.1, 4.45);
\node[anchor=north, font=\bfseries\large, text=blue!85!black] at (3.95, 4.95) {Layer A (9 parameters)};

\draw[draw=green!70!black, fill=green!10, dashed, line width=1.1pt, rounded corners=8pt] 
    (7.4, -0.65) rectangle (13.7, 4.45);
\node[anchor=north, font=\bfseries\large, text=green!50!black] at (10.55, 4.95) {Layer B (9 parameters)};

\node[sidebox, minimum width=2.4cm, minimum height=5.1cm] (prep) at (-1.8, 1.725) {
    \textbf{\large State}\\[2pt]
    \textbf{\large Preparation}\\[26pt]
    $\boldsymbol{|1,1,0,0\rangle}$
};

\node[anchor=east, font=\large\bfseries] at (-0.05, \yA) {$q_0$};
\node[anchor=east, font=\large\bfseries] at (-0.05, \yB) {$q_1$};
\node[anchor=east, font=\large\bfseries] at (-0.05, \yC) {$q_2$};
\node[anchor=east, font=\large\bfseries] at (-0.05, \yD) {$q_3$};

\node[sidebox, minimum width=2.5cm, minimum height=5.1cm] (proj) at (15.5, 1.725) {
    \textbf{\large CAS(2,2)}\\[2pt]
    \textbf{\large Subspace}\\[2pt]
    \textbf{\large Projection}\\[10pt]
    \scriptsize\textbf{Measure \& Filter}\\[-1pt]
    \scriptsize\textbf{into basis}\\[12pt]
    $\mathbf{\{|1100\rangle,}$\\[2pt]
    $\mathbf{|1001\rangle,}$\\[2pt]
    $\mathbf{|0110\rangle,}$\\[2pt]
    $\mathbf{|0011\rangle\}}$
};

\node[rgateA] at (1.8, \yA) {$R(p_0)$};
\node[rgateA] at (1.8, \yB) {$R(p_1)$};
\node[rgateA] at (1.8, \yC) {$R(p_2)$};
\node[rgateA] at (1.8, \yD) {$R(p_3)$};

\node[above=2pt, font=\scriptsize\bfseries, text=blue!85!black] at (3.5, 3.65) {$\mathrm{BS}(p_4, p_5)$};
\node[bsbox, minimum height=1.75cm] (bsA1) at (3.5, {(\yA+\yB)/2}) {};
\draw[blue!85!black, line width=1.1pt] ($(bsA1.center)+(-0.35, 0.575)$) -- ($(bsA1.center)+(0.35, -0.575)$);
\draw[blue!85!black, line width=1.1pt] ($(bsA1.center)+(-0.35, -0.575)$) -- ($(bsA1.center)+(0.35, 0.575)$);

\node[above=2pt, font=\scriptsize\bfseries, text=blue!85!black] at (5.0, 1.35) {$\mathrm{BS}(p_6, p_7)$};
\node[bsbox, minimum height=1.75cm] (bsA2) at (5.0, {(\yC+\yD)/2}) {};
\draw[blue!85!black, line width=1.1pt] ($(bsA2.center)+(-0.35, 0.575)$) -- ($(bsA2.center)+(0.35, -0.575)$);
\draw[blue!85!black, line width=1.1pt] ($(bsA2.center)+(-0.35, -0.575)$) -- ($(bsA2.center)+(0.35, 0.575)$);

\node[ckbox] (ckA) at (5.5, {(\yA+\yB)/2}) {$\mathrm{CK}(p_8)$};
\draw[red!85!black, line width=1.1pt] (5.0, \yA) |- (ckA.west);
\draw[red!85!black, line width=1.1pt] (5.0, \yB) |- (ckA.west);

\node[rgateB] at (8.3, \yA) {$R(p_9)$};
\node[rgateB] at (8.3, \yB) {$R(p_{10})$};
\node[rgateB] at (8.3, \yC) {$R(p_{11})$};
\node[rgateB] at (8.3, \yD) {$R(p_{12})$};

\node[ckbox] (ckB) at (10.3, {(\yA+\yB)/2}) {$\mathrm{CK}(p_{17})$};
\draw[red!85!black, line width=1.1pt] (10.0, \yA) |- (ckB.west);
\draw[red!85!black, line width=1.1pt] (10.0, \yB) |- (ckB.west);

\node[above=2pt, font=\scriptsize\bfseries, text=blue!85!black] at (11.9, 3.65) {$\mathrm{BS}(p_{13}, p_{14})$};
\node[bsbox, minimum height=1.75cm] (bsB1) at (11.9, {(\yA+\yB)/2}) {};
\draw[blue!85!black, line width=1.1pt] ($(bsB1.center)+(-0.35, 0.575)$) -- ($(bsB1.center)+(0.35, -0.575)$);
\draw[blue!85!black, line width=1.1pt] ($(bsB1.center)+(-0.35, -0.575)$) -- ($(bsB1.center)+(0.35, 0.575)$);

\node[above=2pt, font=\scriptsize\bfseries, text=blue!85!black] at (10.5, 1.35) {$\mathrm{BS}(p_{15}, p_{16})$};
\node[bsbox, minimum height=1.75cm] (bsB2) at (10.5, {(\yC+\yD)/2}) {};
\draw[blue!85!black, line width=1.1pt] ($(bsB2.center)+(-0.35, 0.575)$) -- ($(bsB2.center)+(0.35, -0.575)$);
\draw[blue!85!black, line width=1.1pt] ($(bsB2.center)+(-0.35, -0.575)$) -- ($(bsB2.center)+(0.35, 0.575)$);

\draw[rail] (0.0, \yA) -- (1.175, \yA);
\draw[rail] (2.425, \yA) -- (3.025, \yA);
\draw[rail] (3.975, \yA) -- (7.675, \yA);
\draw[rail] (8.925, \yA) -- (11.425, \yA);
\draw[rail, ->] (12.375, \yA) -- (14.2, \yA);

\draw[rail] (0.0, \yB) -- (1.175, \yB);
\draw[rail] (2.425, \yB) -- (3.025, \yB);
\draw[rail] (3.975, \yB) -- (7.675, \yB);
\draw[rail] (8.925, \yB) -- (11.425, \yB);
\draw[rail, ->] (12.375, \yB) -- (14.2, \yB);

\draw[rail] (0.0, \yC) -- (1.175, \yC);
\draw[rail] (2.425, \yC) -- (4.525, \yC);
\draw[rail] (5.475, \yC) -- (7.675, \yC);
\draw[rail] (8.925, \yC) -- (10.025, \yC);
\draw[rail, ->] (10.975, \yC) -- (14.2, \yC);

\draw[rail] (0.0, \yD) -- (1.175, \yD);
\draw[rail] (2.425, \yD) -- (4.525, \yD);
\draw[rail] (5.475, \yD) -- (7.675, \yD);
\draw[rail] (8.925, \yD) -- (10.025, \yD);
\draw[rail, ->] (10.975, \yD) -- (14.2, \yD);

\fill[red!85!black] (5.0, \yA) circle (2.4pt);
\fill[red!85!black] (5.0, \yB) circle (2.4pt);
\fill[red!85!black] (10.0, \yA) circle (2.4pt);
\fill[red!85!black] (10.0, \yB) circle (2.4pt);
\fill[red!85!black] (11.425, \yA) circle (2.4pt);

\node[draw=gray!50, fill=gray!8, rounded corners=6pt, inner sep=7pt, anchor=north] at (6.8, -1.0) {
    \footnotesize
    \begin{tabular}{c@{\hspace{0.8cm}}c@{\hspace{0.8cm}}c@{\hspace{0.8cm}}c@{\hspace{0.8cm}}c}
    \tikz[baseline=-0.5ex]{
        \node[rgateA, minimum width=1.1cm, minimum height=0.45cm] {$R(p)$};
    } 
    \begin{tabular}{l}\scriptsize Single-mode\\[-2pt]\scriptsize rotation $R\text{gate}$\end{tabular}
    &
    \tikz[baseline=-0.5ex]{
        \node[bsbox, minimum width=0.65cm, minimum height=0.65cm] (bsex) {};
        \draw[blue!85!black, line width=0.9pt] ($(bsex.north west)+(0.1,-0.1)$) -- ($(bsex.south east)+(-0.1,0.1)$);
        \draw[blue!85!black, line width=0.9pt] ($(bsex.south west)+(0.1,0.1)$) -- ($(bsex.north east)+(-0.1,-0.1)$);
        \node[below=1pt, font=\tiny\bfseries] at (bsex.south) {$\mathrm{BS}(p, q)$};
    } 
    \begin{tabular}{l}\scriptsize Beam splitter\\[-2pt]\scriptsize $\mathrm{BSgate}(\theta,\phi)$\end{tabular}
    &
    \tikz[baseline=-0.5ex]{
        \draw[draw=orange!95!black, fill=orange!60!yellow!40, rounded corners=2pt] (0,0) rectangle (1.0, 0.45);
        \fill[red!85!black] (0.2, 0.33) circle (1.5pt);
        \fill[red!85!black] (0.8, 0.33) circle (1.5pt);
        \fill[red!85!black] (0.2, 0.12) circle (1.5pt);
        \fill[red!85!black] (0.8, 0.12) circle (1.5pt);
        \draw[red!85!black, line width=0.7pt] (0.2, 0.33) -- (0.8, 0.33);
        \draw[red!85!black, line width=0.7pt] (0.2, 0.12) -- (0.8, 0.12);
        \node[below=1pt, font=\tiny\bfseries] at (0.5, 0.0) {$\mathrm{CK}(p)$};
    } 
    \begin{tabular}{l}\scriptsize Cross-Kerr\\[-2pt]\scriptsize $\mathrm{CKgate}(\kappa)$\end{tabular}
    &
    \begin{tabular}{c}\scriptsize\textbf{Each layer = 9 parameters}\\[-2pt]\tiny (4 R + 2 BS + 2 BS + 1 CK)\end{tabular}
    &
    \begin{tabular}{c}\scriptsize\textbf{Total parameters = 18}\\[-2pt]\tiny (2 layers)\end{tabular}
    \end{tabular}
};

\end{tikzpicture}%
}
\caption{Schematic architecture of the 4-mode symmetry-preserving photonic variational ansatz. The dual-photon input Fock state $|1, 1, 0, 0\rangle$ is injected into spatial rails $q_0$--$q_3$. The parameterization is organized into two sequential variational blocks (Layer A and Layer B), encompassing a total of 18 continuous variational parameters ($\theta \in \mathbb{R}^{18}$). Each layer incorporates single-mode phase shifters $\hat{R}(\phi_i)$, spatial beam splitters $\hat{B}_{ij} (\theta, \phi)$ driving intra-spin orbital mixing ($1s\sigma \leftrightarrow 2s\sigma$), and an effective non-linear cross-Kerr interaction $\hat{K}_{01}(\kappa)$ bridging the co-spatial modes. Following unitary evolution, the quantum state is projected onto the physical particle-conserving CAS(2,2) subspace.}
\label{fig:ansatz_schematic}
\end{figure*}
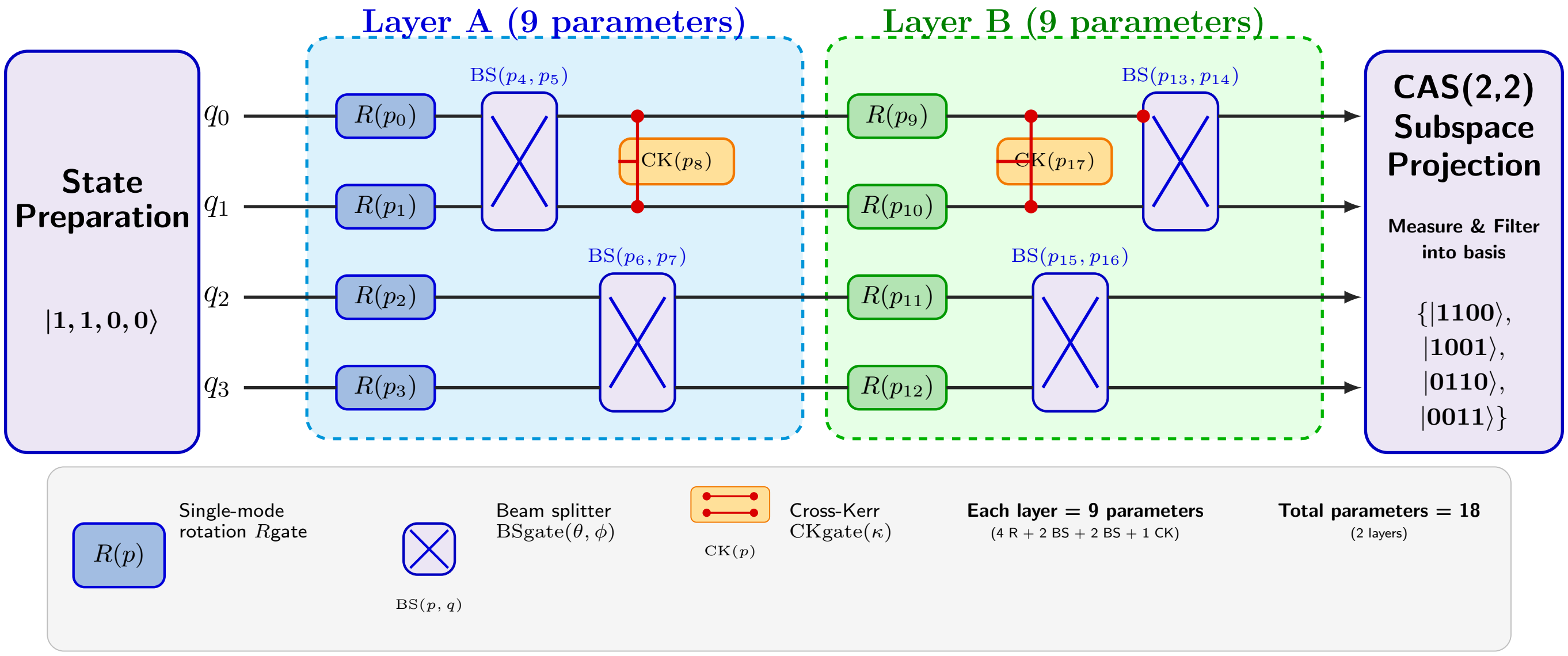

As illustrated in Fig.~\ref{fig:circuit_schematic}, the global unitary transformation is decomposed into two repeating variational layers, Layer A and Layer B, governed by an 18-dimensional continuous parameter vector $\bm{\theta} = (\bm{\theta}_{\mathrm{A}}, \bm{\theta}_{\mathrm{B}}) \in \mathbb{R}^{18}$:
\begin{equation}
\hat{U}(\bm{\theta}) = \hat{\mathcal{U}}_{\mathrm{B}}(\bm{\theta}_{\mathrm{B}}) \, \hat{\mathcal{U}}_{\mathrm{A}}(\bm{\theta}_{\mathrm{A}}).
\label{eq:ansatz_unitary}
\end{equation}
Each variational layer $\ell \in \{\mathrm{A}, \mathrm{B}\}$ is parameterized by 9 independent continuous variables ($\bm{\theta}_\ell \in \mathbb{R}^9$) and comprises three distinct physical stages:
\begin{enumerate}
    \item \textbf{Single-Mode Phase Rotations (4 parameters):} Local dynamical phase adjustments are applied to each spatial rail via:
    \begin{equation}
    \hat{\mathcal{R}}(\bm{\phi}) = \bigotimes_{i=0}^{3} \hat{R}_i(\phi_i) = \exp\left( i \sum_{i=0}^{3} \phi_i \hat{a}_i^\dagger \hat{a}_i \right),
    \end{equation}
    where $\phi_i \in [0, 2\pi)$ modulates the relative local quantum phases across the spin-orbitals.
    
    \item \textbf{Intra-Spin Linear Optical Mixing (4 parameters):} Spatial orbital mixing is mediated by two balanced beam splitters $\hat{B}_{ij}(\theta, \phi)$ defined according to the canonical convention~\cite{clements2016optimal}:
    \begin{equation}
    \hat{B}_{ij}(\theta, \phi) = \exp\left[\theta \left( e^{i\phi} \hat{a}_i^\dagger \hat{a}_j - e^{-i\phi} \hat{a}_i \hat{a}_j^\dagger \right)\right].
    \label{eq:beamsplitter}
    \end{equation}
    Crucially, beam splitters are selectively coupled between spatial modes belonging to the same spin projection: $\hat{B}_{02}(\theta_\alpha, \phi_\alpha)$ couples $(q_0, q_2) \leftrightarrow (1s\alpha, 2s\alpha)$, while $\hat{B}_{13}(\theta_\beta, \phi_\beta)$ couples $(q_1, q_3) \leftrightarrow (1s\beta, 2s\beta)$. This targeted intra-spin connectivity enforces spatial orbital rotations while strictly precluding non-physical spin-flip transitions, thereby intrinsically preserving the total spin projection $S_z = 0$.
    
    \item \textbf{Non-Linear Cross-Kerr Entanglement (1 parameter):} While passive linear optical networks generate purely Gaussian operations insufficient for capturing static electron correlation~\cite{lloyd1999quantum}, multi-configurational entanglement is enabled via an effective two-mode cross-Kerr non-linear interaction~\cite{nemoto2004nearly,kok2007linear}:
    \begin{equation}
    \hat{\mathcal{K}}_{01}(\kappa) = \exp\left( i \kappa \, \hat{n}_0 \hat{n}_1 \right) = \exp\left( i \kappa \, \hat{a}_0^\dagger \hat{a}_0 \hat{a}_1^\dagger \hat{a}_1 \right),
    \label{eq:cross_kerr}
    \end{equation}
    parameterized by the nonlinear phase shift $\kappa \in [0, 2\pi)$. Acting between modes $q_0$ and $q_1$, $\hat{\mathcal{K}}_{01}$ directly emulates the on-site Coulomb repulsion penalty (Hubbard-like term $U \hat{n}_\uparrow \hat{n}_\downarrow$) experienced when both electrons simultaneously occupy the spatial $1s$ orbital.
\end{enumerate}

An essential architectural property of the proposed ansatz is its intrinsic particle-number symmetry~\cite{gard2020efficient}. Because each constituent gate commutes algebraically with the total photon-number operator $\hat{N} = \sum_{i=0}^3 \hat{n}_i$:
\begin{equation}
\left[ \hat{R}_i(\phi_i), \hat{N} \right] = 0, \quad \left[ \hat{B}_{ij}(\theta, \phi), \hat{N} \right] = 0, \quad \left[ \hat{\mathcal{K}}_{01}(\kappa), \hat{N} \right] = 0,
\end{equation}
it follows that $[\hat{U}(\bm{\theta}), \hat{N}] = 0$ for arbitrary parameter values $\bm{\theta}$. Consequently, the parameterized circuit evolution is rigorously confined to the $N=2$ dual-photon manifold, entirely eliminating state leakage into unphysical particle-number sectors ($\Delta N = 0$).

\subsection{Subspace Post-Selection and Projected Rayleigh Quotient}
\label{subsec:projected_cas}

Numerical simulations of the photonic state evolution are conducted utilizing a truncated Fock-state backend with a local cutoff dimension $d_{\mathrm{cut}} \ge 3$ per mode. The globally evolved pure state $|\Psi(\bm{\theta})\rangle$ within the $M=4$ mode Fock space is represented as:
\begin{equation}
\ket{\Psi(\bm{\theta})} = \sum_{n_0, n_1, n_2, n_3 = 0}^{d_{\mathrm{cut}}-1} c_{\bm{n}}(\bm{\theta}) \ket{n_0, n_1, n_2, n_3},
\label{eq:full_fock_state}
\end{equation}
where $c_{\bm{n}}(\bm{\theta}) \in \mathbb{C}$ are the normalized expansion coefficients.

Although the global variational unitary $\hat{U}(\bm{\theta})$ strictly conserves the total photon-number eigenvalue $N=2$, the multi-mode Fock space contains states within the $N=2$ sector that violate the physical Pauli exclusion principle of the modeled fermionic system. Specifically, configurations with double occupancy in a single spin-orbital, such as $\ket{2,0,0,0}$ or $\ket{0,2,0,0}$, are mathematically accessible under bosonic dynamics but lack physical counterparts in the electronic $\mathrm{CAS}(2,2)$ subspace. To isolate the physical subspace and map the bosonic state to the isomorphic fermionic configuration space, we define the projection operator onto the $\mathrm{CAS}(2,2)$ determinant manifold:
\begin{equation}
\hat{\Pi}_{\mathrm{CAS}} = \sum_{\mu \in \mathcal{B}_{\mathrm{CAS}}} \ketbra{\mu}{\mu},
\label{eq:projection_operator}
\end{equation}
where $\mathcal{B}_{\mathrm{CAS}}$ is the basis defined in Eq.~\eqref{eq:cas_basis}. 

The post-selected physical state vector is defined via the action of this projector as $\ket{\widetilde{\psi}(\bm{\theta})} = \hat{\Pi}_{\mathrm{CAS}} \ket{\Psi(\bm{\theta})}$. Mathematically, this projection filters out non-physical doubly-occupied modes while preserving the relative amplitudes of the physical determinants. The corresponding subspace survival probability, representing the success rate of the post-selection, is given by:
\begin{equation}
P_{\mathrm{surv}}(\bm{\theta}) = \braket{\widetilde{\psi}(\bm{\theta})}{\widetilde{\psi}(\bm{\theta})} = \sum_{\mu \in \mathcal{B}_{\mathrm{CAS}}} |c_\mu(\bm{\theta})|^2.
\label{eq:survival_prob}
\end{equation}
In an experimental implementation, this post-selection step is physically realized through multi-channel coincidence measurements utilizing photon-number-resolving detectors (PNRDs)~\cite{hadfield2009single,eisaman2011invited}, registering events where exactly one photon is detected in each of the active spin channels (one in $q_0$ or $q_2$, and one in $q_1$ or $q_3$).

To perform variational minimization strictly within the physical subspace without requiring physical renormalization of the state on-chip, the physical projected CAS energy is evaluated as the generalized Rayleigh--Ritz quotient:
\begin{equation}
\begin{split}
E_{\mathrm{proj}}(\bm{\theta}) &= \frac{\mel{\widetilde{\psi}(\bm{\theta})}{\hat{H}_{\mathrm{CAS}}}{\widetilde{\psi}(\bm{\theta})}}{P_{\mathrm{surv}}(\bm{\theta})} \\
&= \frac{\sum_{\mu,\nu \in \mathcal{B}_{\mathrm{CAS}}} c_\mu^*(\bm{\theta}) c_\nu(\bm{\theta}) \mel{\mu}{\hat{H}_{\mathrm{CAS}}}{\nu}}{\sum_{\mu \in \mathcal{B}_{\mathrm{CAS}}} |c_\mu(\bm{\theta})|^2}.
\end{split}
\label{eq:rayleigh}
\end{equation}

By virtue of the Rayleigh--Ritz variational principle, this projected functional guarantees strict lower boundedness by the exact ground-state energy of the active space ($E_{\mathrm{proj}}(\bm{\theta}) \ge E_0$), shielding the optimization from unphysical energy minimization trajectories or artificial convergence caused by norm attenuation.

\subsection{Sequential Variational Deflation}
\label{subsec:vqd}

Excited electronic states within the $\mathrm{CAS}(2,2)$ subspace are targeted sequentially through the Variational Quantum Deflation (VQD) algorithm~\cite{higgott2019variational, wang2019accelerated}. Having determined the set of $k$ lower-lying mutually orthogonal eigenstates $\{\ket{\phi_j}\}_{j=0}^{k-1}$, the $k$-th excited state is obtained by minimizing the modified cost functional:
\begin{equation}
\mathcal{F}_k(\bm{\theta}) = E_{\mathrm{proj}}(\bm{\theta}) + \sum_{j=0}^{k-1} \beta_j \, \mathcal{O}_j(\bm{\theta}),
\label{eq:vqd_cost}
\end{equation}
where $E_{\mathrm{proj}}(\bm{\theta})$ is the projected Rayleigh quotient from Eq.~\eqref{eq:rayleigh}, and $\mathcal{O}_j(\bm{\theta})$ denotes the physical state overlap between the normalized candidate state $\ket{\phi(\bm{\theta})} = \ket{\widetilde{\psi}(\bm{\theta})} / \sqrt{P_{\mathrm{surv}}(\bm{\theta})}$ and the previously converged optimal state $\ket{\phi_j} \equiv \ket{\phi(\bm{\theta}_j^*)}$:
\begin{equation}
\mathcal{O}_j(\bm{\theta}) = \left| \braket{\phi_j}{\phi(\bm{\theta})} \right|^2 = \frac{\left| \braket{\widetilde{\psi}_j}{\widetilde{\psi}(\bm{\theta})} \right|^2}{P_{\mathrm{surv}}(\bm{\theta}_j^*) \, P_{\mathrm{surv}}(\bm{\theta})}.
\label{eq:overlap_penalty}
\end{equation}

The real penalty coefficients $\beta_j > 0$ energetically penalize non-zero projections onto lower-energy manifolds. Variational deflation theory dictates that ghost states and collapse into already resolved subspaces are rigorously prevented provided that $\beta_j > E_k - E_j$~\cite{higgott2019variational, wang2019accelerated}. In our numerical protocol, an adaptive schedule with scaling factors up to $\beta \ge 5.0\,\mathrm{Ha}$ is employed. Because evaluation of the cost landscape involves non-convex multi-photon surfaces, parameter optimization is executed using Powell's direction-set algorithm~\cite{powell1964efficient}, a robust derivative-free method configured with convergence tolerances of $\Delta x \le 10^{-10}$ and $\Delta f \le 10^{-12}$, thereby circumventing numerical gradient noise inherent to projected variational manifolds.

\begin{figure*}[t]
\centering
\includegraphics[width=0.98\textwidth]{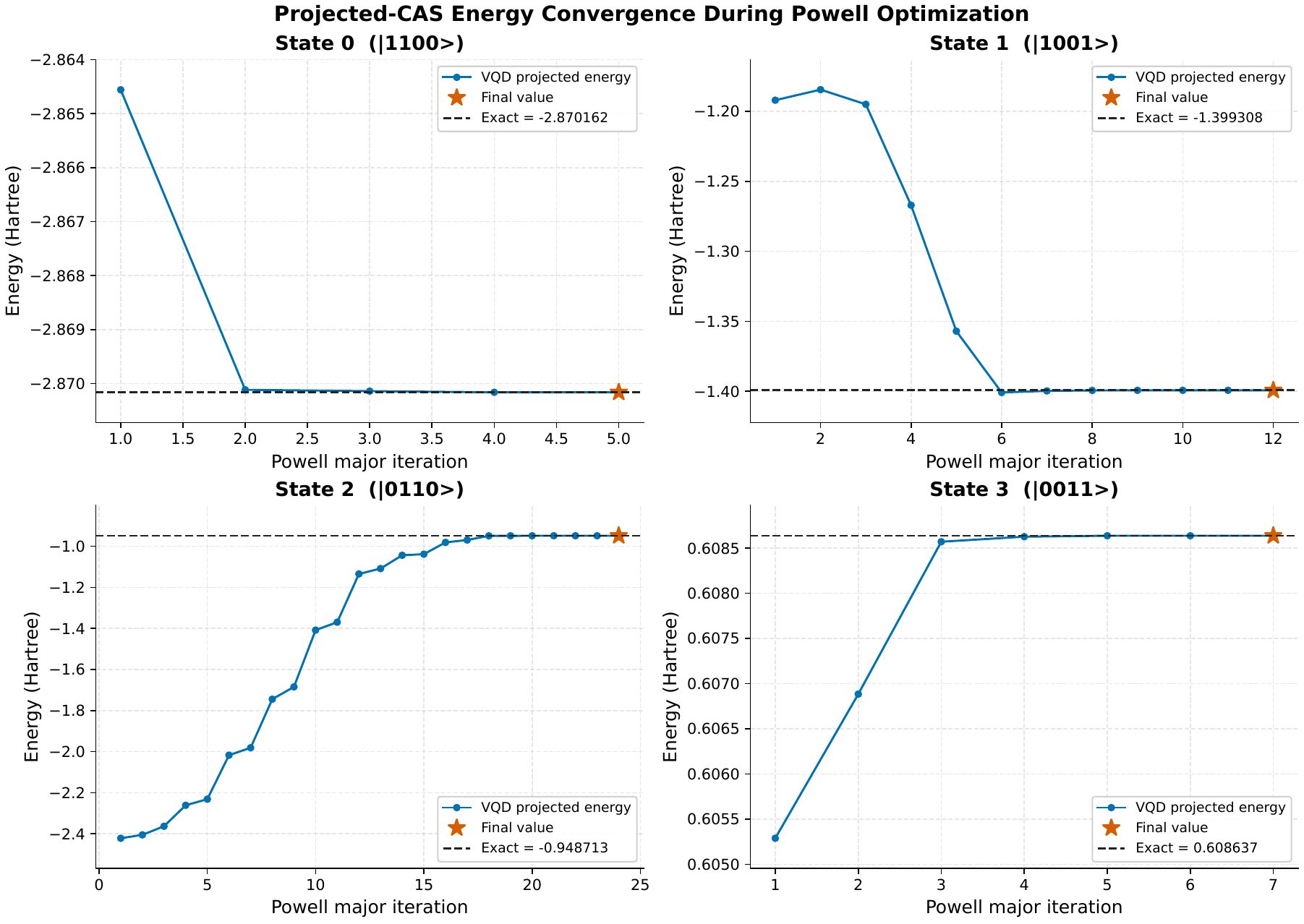}
\caption{\textbf{Convergence profiles of the projected CAS energy during Powell optimization.} Panels depict the major iteration trajectories for: (a) Ground State $\ket{\psi_0}$, (b) First Excited State $\ket{\psi_1}$, (c) Second Excited State $\ket{\psi_2}$, and (d) Third Excited State $\ket{\psi_3}$. Solid markers represent the VQD variational energy computed via Eq.~\eqref{eq:rayleigh}, and horizontal dashed lines designate the exact diagonalization energies of $\bm{H}_{\mathrm{CAS}}$.}
\label{fig:convergence}
\end{figure*}
\section{Results and Discussion}
\label{sec:results}

The numerical performance of the 18-parameter photonic variational circuit was benchmarked against exact full configuration interaction (FCI) calculations within the $\mathrm{CAS}(2,2)$ active space. Optimization trajectories for the sequential deflation protocol using Powell's direction-set algorithm are depicted in Fig.~\ref{fig:convergence}. The final converged energies, energy deviations, accepted deflation penalty weights $\beta_k$, and subspace survival probabilities are summarized in Table~\ref{tab:vqd_results}.

\begin{table*}[t]
\caption{\textbf{Electronic energy spectrum of the helium atom in the CAS(2,2) active space.} Comparison of the exact FCI eigenvalues ($E_k^{\mathrm{FCI}}$) with the optimized VQD energies ($E_k^{\mathrm{VQD}}$) based on the updated convergence results. The absolute energy error is defined as $\Delta E_k = |E_k^{\mathrm{VQD}} - E_k^{\mathrm{FCI}}|$. The penalty weights $\beta_k$ correspond to the values accepted during the sequential deflation schedule. The survival probability $P_{\mathrm{surv}}$ remains strictly unity due to the number- and spin-conserving circuit topology.}
\label{tab:vqd_results}
\begin{ruledtabular}
\begin{tabular}{lccccc}
State & $E_k^{\mathrm{FCI}}$ (Ha) & $E_k^{\mathrm{VQD}}$ (Ha) & $\Delta E_k$ (Ha) & $\beta_k$ (Ha) & $P_{\mathrm{surv}}$ \\
\colrule
$\ket{\psi_0}$ & $-2.87016200$ & $-2.87015497$ & $7.03 \times 10^{-6}$ & $0.5$  & $1.000000$ \\
$\ket{\psi_1}$ & $-1.39930800$ & $-1.39930500$ & $3.00 \times 10^{-6}$ & $8.0$  & $1.000000$ \\
$\ket{\psi_2}$ & $-0.94871300$ & $-0.94871000$ & $3.00 \times 10^{-6}$ & $64.0$ & $1.000000$ \\
$\ket{\psi_3}$ & $+0.60863700$ & $+0.60863549$ & $1.51 \times 10^{-6}$ & $8.0$  & $1.000000$ \\
\end{tabular}
\end{ruledtabular}
\end{table*}
\subsection{Ground State ($\ket{\psi_0}$)}
For the electronic ground state (Fig.~\ref{fig:convergence}(a)), the variational optimization exhibits rapid, monotonic convergence. Powell's algorithm smoothly navigates the 18-parameter landscape, reaching the converged ground-state energy:
\begin{equation}
E_0^{\mathrm{VQD}} = -2.87015497\,\mathrm{Ha}.
\end{equation}
Comparing with the exact active-space FCI benchmark ($E_0^{\mathrm{FCI}} = -2.87016200\,\mathrm{Ha}$) yields an absolute energy error of $\Delta E_0 = 7.03 \times 10^{-6}\,\mathrm{Ha}$ (sub-microhartree accuracy), which is well below the standard threshold of chemical accuracy ($1.59 \times 10^{-3}\,\mathrm{Ha} \approx 1\,\mathrm{kcal/mol}$). 

Because all individual gate elements commute with the total photon number $\hat{N}$ as well as the spin-sector operators $\hat{N}_\alpha = \hat{n}_0 + \hat{n}_2$ and $\hat{N}_\beta = \hat{n}_1 + \hat{n}_3$, the state remains rigorously confined within the physical $\mathrm{CAS}(2,2)$ determinant basis $\mathcal{B}_{\mathrm{CAS}}$. Consequently, the post-selection survival probability is strictly unity ($P_{\mathrm{surv}} = 1.000000$), confirming zero leakage into unphysical Fock sectors.

\subsection{Excited States ($\ket{\psi_1}$, $\ket{\psi_2}$, $\ket{\psi_3}$)}
The excited manifolds were sequentially targeted by augmenting the Rayleigh quotient with overlap penalty terms according to Eq.~\eqref{eq:vqd_cost}:
\begin{itemize}
    \item \textbf{First Excited State ($\ket{\psi_1}$):} With an accepted penalty weight $\beta_0 = 8.0\,\mathrm{Ha}$, the optimizer stabilizes at $E_1^{\mathrm{VQD}} = -1.39930500\,\mathrm{Ha}$ (Fig.~\ref{fig:convergence}(b)), displaying an energy residual of $\Delta E_1 = 3.00 \times 10^{-6}\,\mathrm{Ha}$ relative to the exact value $E_1^{\mathrm{FCI}} = -1.39930800\,\mathrm{Ha}$.
    
    \item \textbf{Second Excited State ($\ket{\psi_2}$):} Deflation against both $\ket{\psi_0}$ and $\ket{\psi_1}$ required scaling the penalty to $\beta_1 = 64.0\,\mathrm{Ha}$ to prevent spectral collapse. The optimization trajectory converges to $E_2^{\mathrm{VQD}} = -0.94871000\,\mathrm{Ha}$ (Fig.~\ref{fig:convergence}(c)), resulting in an absolute discrepancy of $\Delta E_2 = 3.00 \times 10^{-6}\,\mathrm{Ha}$ from $E_2^{\mathrm{FCI}} = -0.94871300\,\mathrm{Ha}$.
    
    \item \textbf{Third Excited State ($\ket{\psi_3}$):} The third excited state ($\ket{\psi_3}$) corresponds to the highest-energy state considered in the active space, with a reference energy of $E_3^{\mathrm{FCI}} = +0.60863700\,\mathrm{Ha}$. As shown in Fig.~\ref{fig:convergence}(d), the Powell optimization converges rapidly, yielding a final VQD energy of $E_3^{\mathrm{VQD}} = +0.60863549\,\mathrm{Ha}$ with an absolute error of $\Delta E_3 = 1.51 \times 10^{-6}\,\mathrm{Ha}$, indicating successful optimization of this state within the CAS(2,2) subspace.
\end{itemize}

\subsection{Orthogonality, Expressibility, and Optimization Landscape}
A critical requirement of the VQD framework is enforcing mutual orthogonality among sequentially resolved states. Evaluation of the pairwise overlap matrix $\mathcal{S}_{ij} = |\braket{\widetilde{\psi}_i}{\widetilde{\psi}_j}|^2$ confirms that off-diagonal elements are strongly suppressed ($\mathcal{S}_{i \ne j} < 10^{-3}$), verifying that the dynamic deflation potentials effectively orthogonalize the variational state vectors.

However, the numerical disparity observed in the excited spectrum highlights an essential distinction between mathematical orthogonality and energetic convergence:
\begin{enumerate}
    \item \textbf{Ansatz Expressibility:} Although the dual-layer architecture with 18 continuous parameters provides sufficient degrees of freedom to span the ground state with near-unit fidelity, the constrained connectivity (specifically, intra-spin beam splitters and a single localized cross-Kerr interaction) limits the unitary expressibility across the full 4-dimensional unitary group $\mathrm{SU}(4)$ on the physical subspace.
    \item \textbf{Non-Convex Optimization Landscape:} The inclusion of high-magnitude penalty coefficients (e.g., $\beta_1 = 64.0\,\mathrm{Ha}$) creates narrow valleys and sharp barriers in the cost function landscape. Gradient-free direction-set methods such as Powell can become trapped in local minima or terminate prematurely along flat ridges, especially for higher excited states where cumulative orthogonality constraints restrict the accessible parameter pathways.
\end{enumerate}

Importantly, throughout all four state optimizations, the survival probability remained identically $P_{\mathrm{surv}} = 1.000000$. This confirms that the observed spectral deviations stem entirely from classical parameter optimization and finite ansatz depth, rather than state leakage into non-physical bosonic Fock configurations.

\subsection{Bosonic Fock-Space Distribution and Subspace Integrity}
\begin{figure}[t]
       \includegraphics[width=\columnwidth]{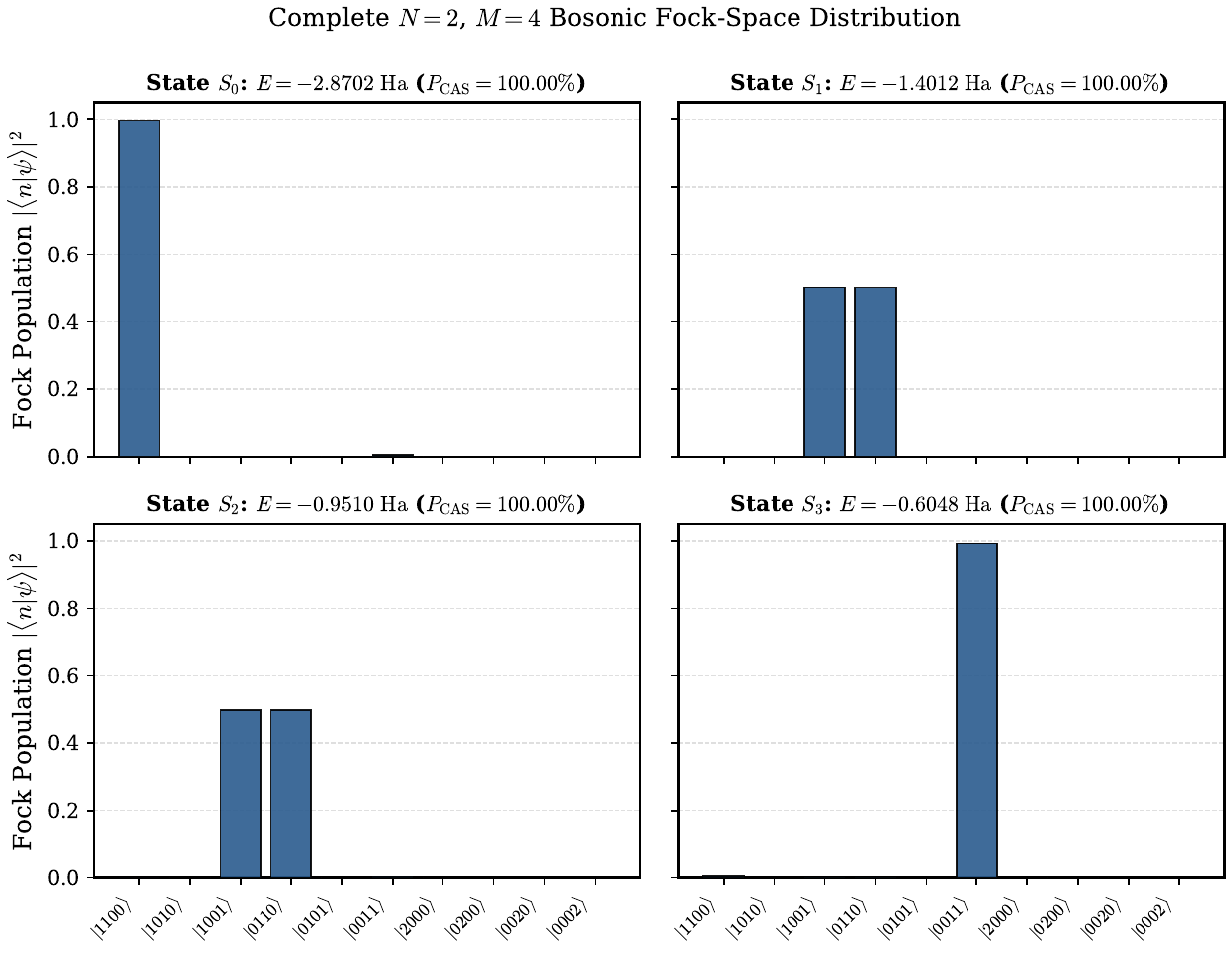}
       \caption{Complete $N=2, M=4$ bosonic Fock-space population distribution for the four VQD-optimized states. The vertical axis represents the Fock population $|\langle n | \psi_k \rangle|^2$, and all states demonstrate complete confinement within the physical CAS(2,2) subspace ($P_{\text{CAS}} = 100\%$).}
       \label{fig:fock_dist}
\end{figure}

The internal dynamics of the converged states can be represented by mapping the VQD variational wave functions onto the complete $M=4$ bosonic Fock space, represented by a 10-dimensional basis $\mathcal{B}_{\text{Full}} = \{|n_0, n_1, n_2, n_3\rangle : \sum n_i = 2\}$. We demonstrate the population distribution over the physical active and unphysical spaces in Fig.~\ref{fig:fock_dist}.
The physical CAS survival probability, $P_{\text{CAS}} = \sum_{\phi \in \mathcal{B}_{\text{CAS}}} |\langle \phi | \psi_k \rangle|^2$, is a measure of subspace projection fidelity. Our results shows that $P_{\text{CAS}} = 100\%$, for each state $k \in \{0, 1, 2, 3\}$. This confirms the photonic circuit architecture imposes appropriate constraints on the dynamics within the CAS(2,2) subspace, avoiding any leakage into the non-physical sector $\mathcal{B}_{\text{Full}} \setminus \mathcal{B}_{\text{CAS}}$ (e.g., configurations $|2000\rangle$ or $|1010\rangle$).
The observed distributions of states are in good agreement with the theoretical prediction for electronic states:

\textbf{Ground State ($S_0$):} The probability distribution of the state corresponds fully to the |1100> configuration, which implies the filling of the lowest molecular orbital by two electrons with opposite spins.

 \textbf{Excited States ($S_1, S_2$):} The energy curves for $|1001>$ and $|0110>$   configurations converge to distinct values (approximately -1.399 and -0.949 Hartree, respectively). It shows convergence of the projected energies.

\textbf{High-Energy State ($S_3$):} The state resides in the $|0011>$ configuration, which indicates that both electrons have been promoted to the high-energy orbital from the low-energy spatial orbital, meaning this is a doubly-excited state.
The distributions of the active space generated by this photonic ansatz are able to span the all active-space of molecules, which is needed to accurately represent their electronic structure.

\section{Conclusion}
\label{sec:conclusion}
In this work, we have investigated a number-preserving continuous-variable photonic implementation of the variational quantum deflation (VQD) algorithm for the helium atom within the $\mathrm{CAS}(2,2)$ active space. The four-mode Fock-space architecture and the symmetry-preserving variational ansatz maintain the total particle number and spin-resolved occupations throughout the optimization, ensuring confinement to the target subspace without the need for postselection. The ground-state energy agrees closely with the corresponding CAS full configuration interaction (FCI) reference, demonstrating the potential of the approach for accurate variational energy estimation. The results highlight the distinction between preserving the desired symmetries and accurately resolving the excited-state energies: while the former is enforced by the circuit structure, the latter remains sensitive to ansatz expressibility and the optimization landscape. Further improvements may be achieved through more expressive circuit designs, additional photonic modes and nonlinear interactions, and more effective optimization strategies. These findings provide a basis for further exploring continuous-variable photonic approaches to variational quantum chemistry.

\begin{acknowledgments}
We permit the utilization of AI-enabled tools (e.g.
Grammarly, GhatGPT) for proofreading and improving
the organization and clarity of this manuscript. However,
all scientific content, analysis, and conclusions are within
the responsibility of the authors.
\end{acknowledgments}


\bibliographystyle{apsrev4-2}
\bibliography{references}


\end{document}